# ESCHERICHIA COLI ACTIVITY CHARACTERIZATION USING A LASER DYNAMIC SPECKLE TECHNIQUE

Evelio E. Ramírez-Miquet†, Luis Martí-López and Orestes R. Contreras-Alarcón.

Centro de Aplicaciones Tecnológicas y Desarrollo Nuclear (CEADEN). Calle 30, No. 502, Miramar, Playa, La Habana, Cuba; ermiquet@ceaden.edu.cu†
†corresponding author

Se presentan los resultados de la aplicación de la técnica de speckle dinámico a la caracterización de actividad bacteriana. La actividad de los patrones de speckle fue detectada en placas Petri de dos compartimientos. Un compartimiento fue inoculado y el otro se utilizó como muestra de control. Las imágenes de speckle fueron procesadas por el método de las diferencias temporales recientemente propuesto. Se realizaron dos pruebas experimentales con tres concentraciones de inóculo diferentes de 0.3, 0.5 y 0.7 McFarland para un total de seis experimentos. Se calcularon las dependencias en el tiempo de la media, la desviación estándar y otros descriptores de la actividad de la evolución de los patrones de speckle correspondientes a la actividad bacteriana y a la actividad del medio de control. En conclusión, la técnica de speckle dinámico propuesta permite caracterizar la actividad de bacterias Escherichia coli en medio de cultivo sólido.

The results of applying a laser dynamic speckle technique to characterize bacterial activity are presented. The speckle activity was detected in two-compartment Petri dishes. One compartment was inoculated and the other one was left as a control blank. The speckled images were processed by the recently reported temporal difference method. Three inoculums of 0.3, 0.5, and 0.7 McFarland units of cell concentration were tested; each inoculum was tested twice for a total of six experiments. The dependences on time of the mean activity, the standard deviation of activity and other descriptors of the speckle pattern evolution were calculated for both the inoculated compartment and the blank. In conclusion the proposed dynamic speckle technique allows characterizing the activity of Escherichia coli bacteria in solid medium.

## INTRODUCTION

Microbiological media such as semen, parasites and bacteria may vary their surface roughness, refractive index, scattering and absorption coefficients and other optical properties in time. If one of such systems is illuminated with coherent light, the speckle pattern of the scattered light also varies in time. The so-called dynamic speckle method exploits the just described causal link to reveal and to follow up such kind of changes. [1-7]

There are several reports on the application of the laser dynamic speckle to characterize biological media dynamics.[4-7] Other methods like infrared thermography,[8-9] microcalorimetry,[10] flow cytometry[11] and optical coherence tomography,[12] have also been employed to describe bacterial activity in liquid and solid media. However, those techniques require expensive equipment and also a well-trained personal to carry out the experimental work. The dynamic speckle technique is simple and practical and the assays based on its principles require minimal resources. Therefore, this technique has many advantages over previously mentioned ones.

To investigate this sort of biological media the recently reported temporal difference method suitable for speckle pattern change detection was employed.[6] In this method, two digitalized speckle images separated by a time interval are subtracted one from the other to detect whether the speckle pattern has changed or not. Therefore, from a sequence of K speckle images a new sequence of K−1 difference images is obtained as shown in Equation 1.

$$D(m,n,k) = |E(m,n,k+1) - E(m,n,k)| \quad (1)$$

Here, E(m,n,k) is the captured irradiance distribution of each image and D(m,n,k) the modulus of the difference of the irradiance of two consecutive images, m and n represent the spatial indices of every pixel and k represents the temporal component of the image sequence.

To describe the evolution in time of the speckle images several descriptors are used.

Equations 2 and 3 show the expressions for the mean ($D_{MEAN}$) and the standard deviation ($D_{STD}$) of D(m,n,k), where N and M are the detector length and width in terms of pixels.

$$D_{MEAN}(k) = \frac{1}{MN} \sum_{n=1}^{N} \sum_{m=1}^{M} D(m,n,k) \quad (2)$$

$$D_{STD}(k) = \left( \frac{\sum_{n=1}^{N} \sum_{m=1}^{M} [D(m,n,k) - D_{MEAN}(m,n,k)]^2}{(N-1)(M-1)} \right)^{1/2} \quad (3)$$

The mean consists on summing the irradiance values and dividing them by the amount of pixels. The $D_{STD}$ has a derivative meaning represented in finite difference form. Using the relation between the derivate and the changes in the speckle pattern, the activity can be characterized.

Another descriptor that can be employed is the cumulative activity $D_{CUM}$ represented in Equation 4. It is a temporal representation of the sum of the mean values of the activity. It permits analyzing the integral of the mean activity in time.

$$D_{CUM}(m,n,k) = \sum_{k=1}^{K} D_{MEAN}(m,n,k) \quad (4)$$

The present work is aimed to a) test a laser dynamic speckle technique combined with the above described temporal difference method for characterizing bacterial activity, b) to verify that the speckle activity of a medium inoculated with bacterial strains clearly differentiates from the activity of a non-inoculated medium, c) to test if different concentrations of the inoculum present different onset times and the phases of growth as Monod[13] defined.

The structure of the work is as follows. In section 2 the setup and the experimental procedure are described. In section 3 the results are analyzed. Afterwards the conclusions, the acknowledgments and the references are presented.

## MATERIALS AND METHODS

The proposed dynamic speckle technique and the temporal difference method were tested for the characterization of Escherichia coli (E. coli) activity in time. This bacteria is typically located inside mammalian intestine and therefore in the sewage, and it is related to some hetero-pathogenic, hetero-toxigenic, hetero-invasive and hemorrhagic diseases.[14-15] Therefore, colony forming unit (CFU) counting of E. coli is used as a contamination index of water and food.[15] Additionally, E. coli bacteria are a good model for studying bacterial growth dynamics under different conditions.

In the experimental series two-compartment Petri dishes were employed. Both compartments were prepared with the same nutrient medium (Mueller Hinton Agar), but E. coli strains were seeded only in one of them. The other one was left as a control blank. Three different inoculum concentrations were tested: 0.3 McFarland, 0.5 McFarland and 0.7 McFarland equivalents to $8.9 \times 10^7$ CFU/mL, $1.5 \times 10^8$ CFU/mL and $2.1 \times 10^8$ CFU/mL of the strain E. coli ATCC (25922). The experiment with each concentration was repeated twice for a total of six experiments. The strains, at different inoculum concentration, were plated using the calibrated swab technique and incubated to give confluent growth, according to the standard procedure.[16]

Figure 1 shows a scheme of the setup. The 90mm two compartment Petri dish (1) is illuminated by the laser diode (2) (Sanyo DL3147-060, wavelength 650 nm, 7 mW output power) which is coupled to a convex lens to form a 40mm × 30mm spot on the Petri dish. The photographic objective (3) (Krasnogorsky factory S. A. Sverev, Industar 61 L/D) has a 50 mm focal distance and is coupled to the detector array (4) of a Vimicro USB CMOS camera ZC0301PLH. The foregoing described device is placed in an incubator (5) that has been stabilized at (37±0.5)ºC, which is reported to be the optimum temperature for E. coli growth to be produced.[17-18] The signal coming from the CMOS camera is introduced in the computer (6) through an USB port. The computer controls the speckle images capture and also process the information using codes programmed in MATLAB™ R2008a.

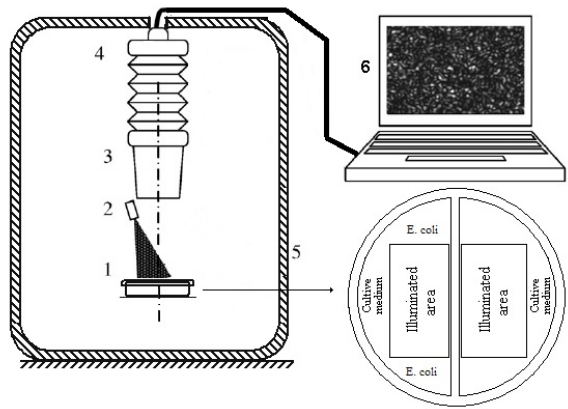

Figure 1. Scheme of the set-up. Petri dish illuminated by the laser light. *E. coli* strains are seeded in one of the compartments.

The experimental procedure can be summarized as follows:

1. The Petri dish with the culture medium is prepared.
2. The incubator and the laser diode are energized. The system reaches the steady state.
3. The strains are seeded in one of the Petri dish compartments.
4. The Petri dish is properly placed in a way that both compartments are illuminated. The CMOS camera is adjusted so that a fully developed speckle pattern can be observed in the computer screen.
5. The incubator is hermetically closed.
6. The captures program is executed and during 24 hours a speckled image is captured every 15 minutes for a total of 96 images.
7. The obtained images are processed.

## RESULTS

Figure 2 shows a 24-hours sequence of speckle images of the Petri dish. It can be clearly observed the evolution of the speckle pattern of the compartment labeled B sown with a strain of E. coli with 0.5 McFarland, while the image of the control compartment, labeled as A, does not present changes of its speckle pattern. The effect of the bacterial activity on the speckle pattern is evident.

Figure 3-4 show the $D_{MEAN}$ and $D_{STD}$ descriptors of the speckle pattern activity for every tested inoculum concentration. The activities for the compartment with strains onset and the culture medium did not onset and its activity values remained quite similar when the experiment was repeated. The cumulative activities in Figure 5 also make evident the difference between the activities of the medium and that one corresponding to the E. coli.

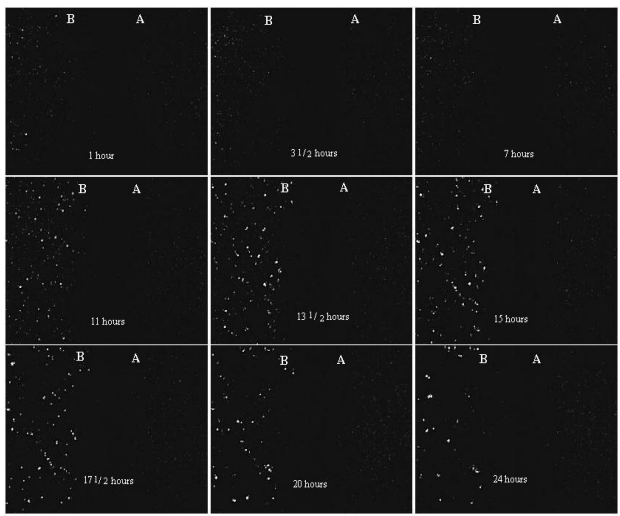

Figure 2. Temporal sequence of the speckle images of a Petri dish where the strains were sown with 0.5 McFarland.

The results obtained with descriptors $D_{MEAN}$ and $D_{STD}$ find analogies with the bacterial growth cycle defined by Monod.[13] Activities represent a lag phase during the first hours. Afterwards, the onset of the E. coli activity takes place and it produces in an exponential way, resembling the so-called exponential phase. A stationary phase comes next and in the end, a death phase can be observed. Figure 6 shows that if the contribution of the solid medium to the cumulative activity is subtracted from the E. coli cumulative activity, it resembles the first three bacterial growth phases.

This means that the bacterial growth has the greater contribution to the activity. However, during the lag phase and the stationary phase in Figures 3-4, the activities values also vary. This occurs due to the contribution the other processes have to the system dynamics. These other processes are the movement inherent to the bacteria, the bacterial metabolism expressed mostly in the excretion and the culture medium water evaporation.

Table I contains the results of the fitting for the exponential phase for every experiment. The activity was fitted in the interval [tonset ; tmax]. The results yielded that an expression like Equation 5 is adequate to describe the behavioral activity of the biological system in this phase.

$$y = e^{a+bt+ct^2} \quad (5)$$

In Equation 5, y represents the activity and t represents time.

Table I
Results for the fitting of activity in the exponential phase.

| Inoculum Concentration (McFarland) | a | b | c | $R^2$ |
|---|---|---|---|---|
| 0.7 Exp 1 | -1.99699±0.5509 | 0.65668±0.1416 | -0.02202±0.0089 | 0,9784 |
| 0.7 Exp 2 | -6.30007±0.8272 | 1.7224±0.19626 | -0.08345±0.0115 | 0.9776 |
| 0.5 Exp 1 | -12.69155±1.1659 | 3.37399±0.2873 | -0.18763±0.0175 | 0.9845 |
| 0.5 Exp 2 | -13.23182±0.7695 | 3.81738±0.21346 | -0.23278±0.0147 | 0,9968 |
| 0.3 Exp 1 | -4.97215±0.7343 | 1.43203±0.18421 | -0.07276±0.0114 | 0.9830 |
| 0.3 Exp 2 | -7.66466±0.8647 | 1.89003±0.2022 | -0.08881±0.0117 | 0.9904 |

The criterion employed for the determination of the onset time of the bacterial activity is based on the following consideration. The medium activity is subtracted from bacterial activity. When activity values increase and the relative difference between two consecutive values is higher than 10% then it will be considered that the activity due to bacteria has onset.

Table II contains the results of the onset time for every experiment with different inoculum concentrations.

Table II
Results for the fitting of activity in the exponential phase.

| Concentration | Exp | Onset time | Mean Activity | Relative Difference |
|---|---|---|---|---|
| 0.7 McFarland | 1 | 4.75 hours | 1.2408 a.u | 11% |
| 0.7 McFarland | 2 | 4.75 hours | 1.1114 a.u | 15% |
| 0.5 McFarland | 1 | 5.75 hours | 1.1346 a.u | 18% |
| 0.5 McFarland | 2 | 5. 5 hours | 1.5109 a.u | 39% |
| 0.3 McFarland | 1 | 6 hours | 1.6375 a.u | 41% |
| 0.3 McFarland | 2 | 6.25 hours | 1.3830 a.u | 23% |

The onset time for the higher and medium concentration are similar to those reported by Hans et al.[8] and Hernández-Eugenio et al.[9] using thermal images. They detected E. coli presence in five hours. Therefore the dynamic speckle based method is able to obtain the same result using a low-cost configuration.

An order of difference in the amount of cells that were sown was significant to the onset time. The difference in the activity onset time for those experiments using 0.3 and 0.7 McFarland was almost two hours.

These experimental results make evident that the laser dynamic speckle technique detects E. coli activity and the temporal difference method permits its characterization.

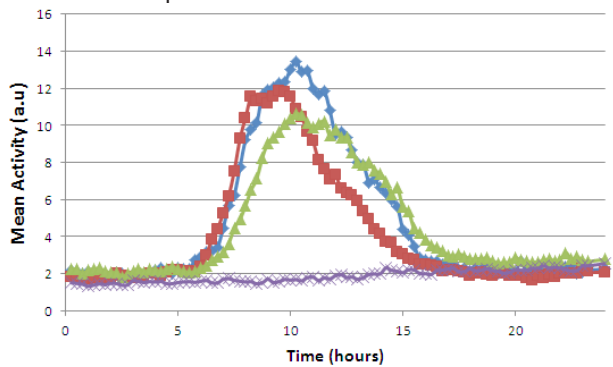

Figure 3. Mean Activities for 0.7 McFarland (blue), for 0.5 McFarland (red) and for 0.3 McFarland (green) and for Culture Medium (violet).

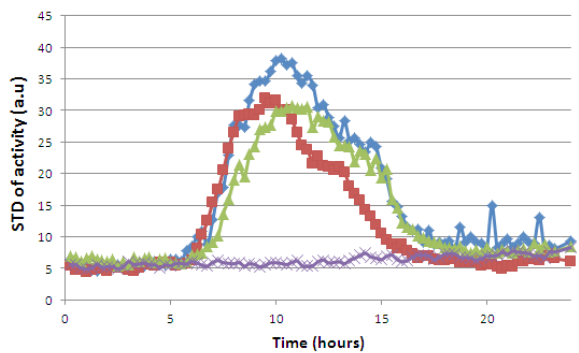

Figure 4. STD Activities for 0.7 McFarland (blue), for 0.5 McFarland (red) and for 0.3 McFarland (green) and for Culture Medium (violet).

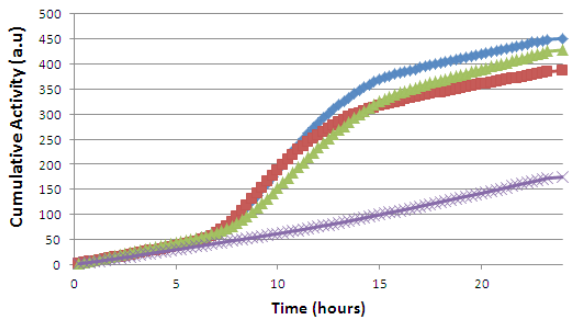

Figure 5. Cumulative Activities for 0.7 McFarland (blue), for 0.5 McFarland (red) for 0.3 McFarland (green) and for Culture Medium (violet).

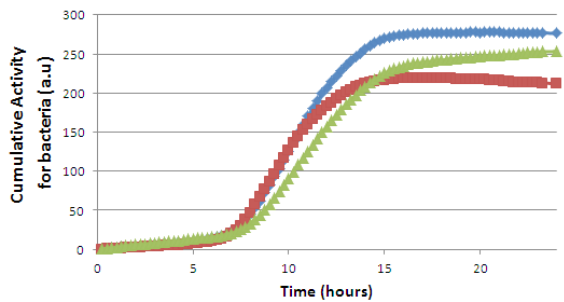

Figure 6. Cumulative Activities due to bacteria for 0.7 McFarland (blue), for 0.5 McFarland (red) and for 0.3 McFarland (green).

## CONCLUSIONS

The dynamic speckle method combined with the temporal difference processing permits the characterization of the E. coli activity in Petri dishes. It not only differentiate the activity of the inoculated compartment from the non-inoculated one but also was able to detect differences in the onset time of bacteria growth for three inoculums of 0.3 McFarland, 0.5 McFarland and 0.7 McFarland. Furthermore, the speckle descriptors $D_{MEAN}$ and $D_{STD}$ presented analogies with all bacterial growth phases, while the accumulated activity $D_{CUM}$ reproduces three of them.

## ACKNOWLEDGMENTS

This work was carried out in the frame of the project "Fase" (Phase) of AENTA (CITMA, Cuba) with support of CEADEN. Special thanks to the staff of the DIRAMIC Microbiology Laboratory at CNIC for support and collaboration. Also thanks to Dr. Luz Marina Miquet, Dr. Efrén Andrades, Lic. Grether Barrera and Dr. Dinorah Hernández for their help.